\documentstyle[12pt]{article}
\newcommand{\be}{\begin{equation}}
\newcommand{\ee}{\end{equation}}
\newcommand{\bea}{\begin{eqnarray}}
\newcommand{\eea}{\end{eqnarray}}
\newcommand{\ba}{\begin{array}{l}}
\newcommand{\ea}{\end{array}}
\newcommand{\mufv}{\gamma^{\mu}\gamma^5}

\newcommand{\gf}{\gamma^5}
\newcommand{\bb}{\begin{thebibliography}}
\newcommand{\eb}{\end{thebibliography}}
\newcommand{\ci}[1]{\cite{#1}}
\newcommand{\lab}[1]{\label{#1}}
\newcommand{\re}[1]{(\ref{#1})}
\newcommand{\Tr}{\mbox{Tr\,}}
\newcommand{\Ds}{\displaystyle}

\title{
\hfill  {\small  Report-no: U.of Alberta Thy 05-95,
hep-th/9502139}  \vskip 0.5cm
ON THE HIGHER-LOOP CONTRIBUTIONS TO THE AXIAL ANOMALY }
\author{B.A.FAIZULLAEV\\
 Theoretical Physics Department, Tashkent State University, \\
 Tashkent 700095, Uzbekistan \\
M.M.MUSAKHANOV\thanks{Permanent address: Theoretical
                              Physics Department, Tashkent State
                                University, Tashkent 700095,
              Uzbekistan, e-mail: yousuf@iceeat.silk.glas.apc.org} \\
Department  of Physics, University of  Alberta, \\
Edmonton, Canada  T6G 2J1 \\
and \\ TRIUMF, 4004 Wesbrook Mall, \\
Vancouver, British Columbia,
Canada, V6T 2A3, \\
e-mail: yousuf@phys.ualberta.ca \\
                        and\\
N.K.PAK\\
 Physics Department, Middle East Technical University \\
     Ankara 06531, Turkey}

\begin{document}
\begin{titlepage}
%\medskip
%\rightline {Report-no: Alberta Thy 05-95 }
%\medskip
\maketitle

%\rightline {Report-no: Alberta Thy 05-95 }

\begin{abstract}

The problem of the higher-loop contributions to the
axial anomaly is reexamined by a new method.
We demonstrate that these contributions
depend on the order of the calculations. If
the divergence of the axial current by nonperturbative Fujikawa
method is calculated first and then average it
over the photon field in the presence of an external photon source,
a nonzero contribution is obtained.

However perturbative Feynman diagram method  has an uncertainty.
Depending on the
order of the calculations above mentioned or zero  results are
obtained.
\end{abstract}
\end{titlepage}
\section{ Introduction}
The problem of the axial anomaly \ci{1,2} is one of the
attractive problems of the quantum field theory.
In the framework of QED it was first discussed in \ci{1,3},
using Feynman diagram  method,
 with the conclusion that two and higher-loop contribution to the
divergence of the axial current is zero.

On other hand in \ci{4} this anomaly was calculated by a
nonperturbative
technique, based on the observation that the anomaly is due to the
noninvariance
of the fermion measure under axial transformations of the fermion
fields.

In a recent work \ci{5},  the 3-loop contribution to
the divergence of the axial current was shown to be nonzero.
This conclusion was supported in \ci{6} based on
the investigations of the renormalization properties of axial current
and
its divergence.

Given the controversy we would like to discuss this problem by
using yet another method \ci{7}.
This method is based on an old formula of DeWitt \ci{8}
that connects the vacuum expectation of a quantum functional with the
classical
ones  ( and its derivatives with respect to the classical fields ).
Using this method we have previously calculated the two - loop
effective action for scalar
$\lambda\varphi^4$ theory  and for spinor electrodynamics in
\ci{7,8}.

In this paper we discuss the divergence of the axial current in
the simplest model which describes interacting massless fermion
$\psi$
and photon fields $A^{\mu}$
via axial-vector coupling. The Lagrangian of this model
is invariant under local axial transformations
and have the form (with gauge fixing):
\be
L=-\frac{1}{4}F_{\mu\nu}F^{\mu\nu}+\bar\psi \left[i\left(\hat
\partial - ie\hat A\gf\right ) \right]\psi -\frac{1}{2\alpha}\left
 (f(\partial^2)\partial_{\mu}A^{\mu}\right )^2.
\lab{lag}
\ee
The regularization is carried out by adding
to the Lagrangian \re{lag} the higher derivative term \ci{9}:
\be
\frac{1}{4}F_{\mu\nu}\frac{\partial^2}{\Lambda^2}F^{\mu\nu},
\lab{reg}
\ee
which regularizes the free photon propagator. For consistency
we choose the function $f(\partial^2) $ as
$$f(\partial^2)=\partial^2 + \kappa^2, $$
where $\kappa^2 $ is an arbitrary parameter.
Then the final form of the model we consider is:
\be
L=-\frac{1}{4}F_{\mu\nu}F^{\mu\nu}+
\frac{1}{4}F_{\mu\nu}\frac{\partial^2}{\Lambda^2}F^{\mu\nu}
+\bar\psi \left[i\left(\hat\partial - ie\hat A\gf\right )
\right]\psi
 -\frac{1}{2\alpha}\left(f(\partial^2)\partial_{\mu}A^{\mu}\right
)^2.
\lab{e1}
\ee

\section{DeWitt Method}
In this section we would like to review briefly the DeWitt method
which we will need in the following discussion.

DeWitt's formula \ci{8} relates the vacuum expectation
value of any quantum functional $ Q[\hat\varphi_i] $, in the
presence of the source $ J $,
to the classical one - $Q[\varphi_i] $ :
\be
\langle Q[\hat \varphi]\rangle
 =:\exp  \left( \frac {i}{\hbar}\sum_{n=2}^{\infty }
 \frac{(-i\hbar )^n}{n!}
 G^{i_1\cdots i_n}
\frac {\delta^n}{\delta \varphi^{i_1}\cdots \delta \varphi^{i_n}}
\right):
 Q[\varphi], \lab{e2}
\ee
where $\hat\varphi_i=\hat\varphi(x_i) $ denotes any quantum field
and
$\varphi_i=\delta W/\delta J_i $ is its vacuum expectation value,
$W=-i\log Z[J] $ is the generating functional of the connected
Green's functions, and
$$G^{i_1\cdots i_n}=\frac{\delta^n W}{\delta J_{i_1}\cdots \delta
J_{i_n}}$$
is the connected $n $-point Green's functions. Note that repeated
indices
imply a summation over discrete
indices and integration over the continuous ones.
The colons in Eq.\re{e2} mean that derivatives in exponent acts
only on functional  $Q[\varphi] $.

Thus  the vacuum expectation value of any quantum
functional
is defined  by the corresponding classical functional and the
connected total
Green's functions.

Let us define the generating functional
\be
 Z[J_\mu]=\exp\{iW[J_\mu]\}
= \int D\hat{A_{\mu}} D\hat{\psi}D\hat{\bar\psi}
\exp\left(i\int\left(L+J^{\mu}\hat{A_{\mu}} \right)dx\right).
\lab{e4}
\ee
The integrations over spinor variables in Eq.\re{e4} leads to:
\be Z[J_\mu]= \int{D\hat A_\mu}\exp\left({i}S_{eff}+i\int J^{\mu}
\hat A_{\mu}dx\right),\ee
where
\be S_{eff}=\frac{1}{2}\hat A_\mu K^{-1}_{\mu\nu}A_\nu
+i\hbar \Tr\log\hat K, \ee
and
\be
\Ds{
K^{\mu\nu}=\frac{1}{\partial^4\left(1-\partial^2\Lambda^{-2}\right)}
\left(g^{\mu\nu}\partial^2-\partial^{\mu}\partial^{\nu}\right)+
\frac{\alpha}{f^2}\frac{\partial^{\mu}\partial^{\nu}}{\partial^4},
\qquad
\hat K^{-1}=i\hat\partial+e\hat A\gf.}
\ee
We next define the effective action
\be \Gamma[A_\mu]=W[J_\mu]-J_\mu A^\mu. \ee
According to \ci{8} and \ci{7}:
\be
\frac{\delta \Gamma}{\delta A_\mu}=\langle \frac{\delta
S_{eff}}{\delta \hat
A_\mu}\rangle,
\ee
and from Eq.\re{e2} we have the following equation:
\be
\frac{\delta\Gamma}{\delta A_{\mu}(x)}= :\exp(\hat G):\frac{\delta
S_{eff}}
{\delta A_{\mu}(x)},\ee
where
\be\hat G=\frac{-i\hbar}{2}D^{\mu\nu}(y,z)\frac{\delta^2}{\delta
A^\mu(y)\delta A^\nu(z)}
-\frac{\hbar^2}{6} D^{\mu\nu\lambda}(y,z,t)\frac{\delta^3}{\delta
A^\mu(y)\delta A^\nu(z)\delta A^\lambda(t)}
+\ldots,\ee
and
$$D^{\mu\nu\ldots\lambda}(x,y,\ldots,t)= \frac{\delta^n W}{\delta
A_\mu(y)\delta A_\nu(z)\ldots\delta A_\lambda(t)}$$
are connected Green's functions of the vector fields $A_{\mu}$.

\section{The Divergence of the Axial Current}

Using the formalism developed in Sec.2 it is easy to show that
\be
\ba
 \langle e\hat{\bar\psi}(x)\gamma^{\mu}\gamma_{5}
\hat{\psi}(x)\rangle_{J}=\delta\Gamma/\delta A_\mu(x)
-K^{-1\mu\nu}(x,y)A_{\nu}(y)=\\ \\
=-i\hbar e\Tr\left(\hat K(x,x)\mufv\right)-
\Ds{\frac{\hbar^2}{2}e D^{\alpha\beta}(y,z)\frac{\delta^2}{\delta
A_{\alpha}(y)
\delta A_{\beta}(z)}\Tr\left(\hat K(x,x)\mufv\right) + \cdots.}
\ea
\ee
The dots denote the terms with higher order derivatives over
$A_{\mu}$.
The left side of this equation is vacuum expectation of the axial
current
$j^{5}_{\mu}$ in the presence of the source $J$.
The vacuum expectation of the
divergence of the axial current in the presence of the source $J$
is:
\be \ba
 \langle \partial_{\mu} j^{5}_{\mu}
\rangle_{J} =-ie\hbar\partial^x_{\mu}\Tr\left(\hat
K(x,x)\mufv\right)
- \\ \\
-\Ds{\frac{\hbar^2}{2}e\partial^x_{\mu}
D^{\alpha\beta}(y,z)\frac{\delta^2}{\delta A_{\alpha}(y)
\delta A_{\beta}(z)}\Tr\left(\hat K(x,x)\mufv\right)+\cdots}.
\lab{a1}
\ea
\ee
In the framework of the model Eq.\re{lag}
% the one-loop contribution to
the
divergence of the axial current was calculated by the nonperturbative
method in \ci{4},  and by perturbative one in \ci{9} (for
other models see e.g. \ci{1}, \ci{3}).
Let us first  consider the nonperturbative method  of Fujikawa \ci{4}
to
calculate the divergence of the axial current.
Following  this method we have as a result of the noninvariance of
the
fermion
measure  in Eq.\re{e4} (we keep the conservation of the vector
current):
 \be
\ba
\langle \partial_{\mu} j^{5}_{\mu}\rangle_{J} ^{NP}
=\int D\hat{ A_\mu}D\hat{\psi} D\hat{\bar\psi}
e\partial_{\mu} ^{x}(\hat{\bar\psi}(x)\gamma^{\mu}\gamma_{5}
\hat{\psi}(x))
\exp\left(i\int\left(L+J^{\mu}\hat{A_{\mu}} \right)dx\right) \\ \\
\Ds{=\int{D\hat{A_\mu}}
\hbar\frac{e^2}{16\pi^2}\epsilon^{\mu\nu\lambda\sigma}
\hat F_{\mu\nu}(x) \hat F_{\lambda\sigma}(x)
\exp\left({i}S_{eff}+i\int J^{\mu}\hat{A}_{\mu}dx\right ) .}
\ea
\lab{fujikawa}
\ee
Application of the DeWitt's formula \re{e2} to the right hand side of
the
last equation gives
\be
\langle\partial_\mu j^{\mu}_5(x)\rangle _{J}^{NP} =
\hbar\frac{e^2}{16\pi^2}
\epsilon^{\mu\nu\lambda\sigma}
F_{\mu\nu}(x) F_{\lambda\sigma}(x)
%\tilde F_{\mu\nu}F^{\mu\nu}
- i\hbar^2\frac{e^3}{2\pi^2}\epsilon_{\mu\nu\lambda\sigma}
\partial^{\mu}_z\partial^{\lambda}_y D^{\nu\sigma}(y,z)|_{y=z=x}.
\lab{a2} \ee
This equation is the main result of this paper.
Now implication of this result will be discussed.

We would like to stress that, to find this result we
first calculate the path integral over fermions with further
formal calculations of the path integral over bosons.
This prescription completely coincides with the suggestion of
F.Berezin \ci{berezin}. He gave the order of calculations of the
path integral over fermions and bosons as a definition: first
integrate over fermions, and next over bosons.

By comparison with Eq.\re{a1} we get
\be
-ie\hbar \partial_{\mu}
\Tr\left(\hat{K}(x,x)\mufv\right)=\hbar\frac{e^2}{16\pi^2}
\epsilon^{\mu\nu\lambda\sigma}
F_{\mu\nu}(x) F_{\lambda\sigma}(x).
\lab{e10}
\ee  \\

Now we discuss the calculation \ci{3}, \ci{9} of the same quantity
by
perturbative method.
The left side of Eq.\re{e10} in fact is the one-loop fermion
contribution.
In order to calculate this contribution we need additional (to Eq.
\re{reg})
regularization of this contribution. We can choose Pauli--Villars
method for this one to preserve the conservation of vector currents,
and get the same answer as the right hand side of Eq.\re{e10}.

As we can see from Eq.\re{a1} to compute the higher-loop
contributions we must perform three operations: the divergence,
the functional derivatives, and the trace over discrete and
continuous coordinates.
The final results  depend on the order of these calculations.

Calculated the variational derivatives on $A_{\mu}$ first and
then the divergence and the trace  we get for the
second term on the r.h.s of  Eq.\re{a1}:
\be
\ba
\Ds{-\frac{\hbar^2}{2}e \Tr
D^{\alpha\beta}(y,z)\frac{\delta^2}{\delta A_{\alpha}(y)
\delta A_{\beta}(z)}\left(\hat K(x,x)\mufv\right)}=\\ \\
\Ds{-\frac{e^3\hbar^2}{2}\Tr D^{\alpha\beta}(y,z)\left(\mufv\left(
\hat K(x,y)\gamma_{\alpha}\gamma_5
\hat K(y,z)\gamma_{\beta}\gamma_5\hat
K(z,x)+(y,\alpha\leftrightarrow
z,\beta)\right)\right)}.
\ea
\lab{per}
\ee
Note that in this case the Pauli-Villars procedure is not needed
since
the convergence of the integrals in Eq.\re{per}
is insured by the regularization prescription (see Eq.\re{reg}) and
the traces can be computed.
Acting wiht the operator $\partial^x_{\mu}$ on this expression we get
for the
trace
part of this term:
\be
\ba
\Ds{\Tr\left[\gf\delta (y-x)
\left(\hat K(z,y)\gamma_{\beta}\hat K(x,z)\gamma_{\alpha}-
\hat K(y,z)\gamma_{\alpha}\hat K(z,x)\gamma_{\beta}\right)\right]+}\\
\\
\Ds{\Tr\left[\gf\delta(x-z)
\left(\hat K(y,z)\gamma_{\alpha}\hat K(x,y)\gamma_{\beta}-
\hat K(z,y)\gamma_{\beta}\hat
K(y,x)\gamma_{\alpha}\right)\right]=0}.
\lab{e19}
\ea
\ee
That is, this contribution to the anomaly vanishes.

Proceeding similarly  we will have zero contribution to the
divergence of the
axial
current from the other terms in Eq.\re{a1} as well.

Next, we would like to change the order of
operations.
Acting by $\partial^x_{\mu} $ first  and then calculating the
trace over fermions (which would require Pauli-Villars
regularization), and noting
that r.h.s. of Eq.\re{e10} is quadratic in $A_\alpha(x) $,
our series \re{a1} can be reduced to the same final form
of the nonperturbative method given in Eq.\re{a2}.
%:
%\be
%\langle\partial_\mu j^{\mu}_5(x)\rangle=
%\hbar\frac{e^2}{16\pi^2}\tilde F_{\mu\nu}F^{\mu\nu} -
%i\hbar^2\frac{e^3}{2\pi^2}\epsilon_{\mu\nu\lambda\sigma}
%\partial^{\mu}_z\partial^{\lambda}_y D^{\nu\sigma}(y,z)|_{y=z=x}.
%\lab{a2} \ee
Since $D^{\nu\sigma}(y,z) $ is the full
connected Green's function of the photon,
the second term of this expression contains all the higher-loop
radiative corrections.

It is clear that the second term in Eq.\re{a2} is at
least
third order in $\hbar $, with the substitution
of the Dyson equation for $D^{\nu\sigma}(y,z) $ into Eq.\re{a2} (see
Appendix).
Indeed for the second term in Eq. \re{a2} we have
\be
\ba
\Ds{i\hbar^2\frac{e^3}{2\pi^2}\epsilon_{\mu\nu\lambda\sigma}
\partial^{\mu}_z\partial^{\lambda}_y\left[-K^{\nu\sigma}(y,z)-
ie\hbar\Tr\left(\gamma_{\rho}\gf G^{\nu}(t,y,t)
K^{\rho\sigma}(t,z)\right)\right]=}
\\ \\
\Ds{=\frac{e^4\hbar^3}{2\pi^2}\epsilon_{\mu\nu\lambda\sigma}
\Tr\left(\gamma^{\nu}\gf\partial^{\lambda}_y
G^{\sigma}(t,y,t)\right)
\frac{1}{\partial^2}\partial^{\mu}_z\delta(t-z)}.
\ea
\ee
Here we used the fact that the free Green's function of photon
$K^{\mu\nu}(x,y) $ is symmetric
in $\mu $ and $\nu $. After some algebra, for
three-loop
contribution we get
\be \ba
\Ds{\frac{e^5\hbar^3}{2\pi^2}\epsilon_{\mu\nu\lambda\sigma}
\Tr\left(\gamma^{\nu}\hat K(t,v)\gamma^{\sigma}\hat K(v,t)\right)
\frac{1}{\partial^2\left(1-\partial^2/\Lambda^{2}\right)}
\partial^{\mu}_z\delta(t-z)\cdot} \\ \\
\Ds{\cdot\frac{1}{\partial^2\left(1-\partial^2/\Lambda^{2}\right)}
\partial^{\lambda}_y\delta(y-v)|_{y=z=x}.}
\ea
\lab{3-loop}
 \ee
We can conclude that
the  nonperturbative method of Fujikawa  provides us with a well
defined
procedure:
\begin{itemize}
\item  At first, we  calculate the divergence of the axial current in
the presence
of  the photon field by regularization of  the measure of the path
integral over fermions to get Eq. \re{e10}(We preserve the
conservation of the vector current).
Then the result  of this calculation
is  exact. (It is important that  this result
coincides with the result of the perturbative calculations of
Feynman
fermion triangle  diagram via Pauli-Villars regularization
prescription);
\item  Next,  we calculate the functional derivatives over
the photon fields,
in Eq.\re{a1} to get our final result Eq.\re{a2}.
\end{itemize}

On the other hand,  perturbative calculations  of
Feynman diagrams, in principle, permit a change of the order of the
calculations over fermion and boson variables in higher-loop diagrams
as in Eq.\re{3-loop}.
As a result we  can have zero contribution  as is demonstrated in Eq.
\re{e19}.

\section*{Acknowledgements}
Two of the authors (B.F. and M.M.) would like to thank  B.L.Ioffe and
L.B.Okun for the useful discussions during the First Turkish-Russian
Summer School in Physics at TUBITAK, Turkey. Also, one of the authors
(M.M.) would like to thank F.Khanna for the useful discussions of the
revised version of the paper and the warm hospitality.
This work was supported in part by the
 Uzbekistan State Committee for Science and Technique, International
Science Foundation and INTAS Grants.

\section*{Appendix}
In \ci{7} the system of equations for the effective
action in QED was derived.
In the present case the first equation of this system is
(we will present it in the form where $\eta=\bar\eta=0$):
$$\frac{\delta\Gamma}{\delta A_{\mu}}=K^{-1\mu\nu}A_{\nu}+
ie\hbar\Tr\left(\mufv G(x,x)\right).$$
Differentiating this equation with respect to $A_{\nu} $ and using
the identity
$$ D_{\mu\nu}\frac{\delta^2 \Gamma}{\delta A_{\nu}\delta
A_{\lambda}}=-\delta^\lambda_\mu$$
we get the following Dyson equation for total Green's function
$D_{\mu\nu}$:
$$D^{\mu\nu}(x,y)=-K^{\mu\nu}(x,y) -
ie\hbar\Tr\left(\gamma_{\lambda}\gf
G^{\mu}(z,x,z)\right)K^{\lambda\nu}(z,y), $$
where $ G^{\mu}(z,x,z)$ is three-point connected Green's function
which is related
to the three-point total vertex function $\Gamma^{\mu}(x,y,z)$ as:
$$ \hat G^{\nu}(x,y,z)= \hat
G(x,x_1)\Gamma^{\lambda}(x_1,y_1,z_1)\hat
G(z_1,z)D^{\nu\lambda}(y_1,y), $$
where $ \hat G(x,y) $ is the full Green's function of the fermion.


\begin{thebibliography}{99}
\bibitem{1} S.Adler, {\it Phys.Rev.} {\bf 177}(1969)2426.
\bibitem{2} J.S.Bell and R.Jackiw,{\it Nuovo Cimento} {\bf
60}(1969)47.
\bibitem{3} S.Adler and W.Bardeen, {\it Phys.Rev.} {\bf
182}(1969)1517.
\bibitem{4} K.Fujikawa, {\it Phys.Rev.}{\bf D21}(1980)2848;
ibid.{\bf D29}(1984)285; \\
M.B.Einhorn and D.R.T.Jones, {\it Phys.Rev.}{\bf D29}(1984)331.
\bibitem{5} A.A.Anselm and A.A.Iogansen, {\it JETP Letters} {\bf
49}(1989)185.
\bibitem{6} A.V.Efremov and O.V.Teryaev, {\it Yad.Fiz.}{\bf
51}(1992)1492.
\bibitem{7} B.A.Faizullaev and M.M.Musakhanov, {\it Turk.J.of
                                               Phys.}{\bf 17}(1993)
717;\\
            B.A.Faizullaev and M.M.Musakhanov, {\it
Annals of Phys.}(NY) (in press).
\bibitem{8} B.S.DeWitt, {\it Dynamical Theory of Groups and Fields}
                       (Gordon and Breach, N.-Y.) (1965) .
\bibitem{9} A.Slavnov and L.Faddeev, {\it Gauge Fields. An
Introduction
                                  to Quantum Theory} (Benjamin,
N.-Y.)1980.
\bibitem{10} R.Jackiw, in {\it Lectures on Current Algebra and its
Applications},
ed. by S.Treiman et al. (Princeton University Press, Princeton,
N.J.,1972).
\bibitem{berezin} F.A.Berezin, {\it Method of Second Quantization}
(in russian) (Nauka, Moscow) 1986.
\end{thebibliography}
\end{document}